\documentclass[aps,prapplied,reprint,showkeys,floatfix]{revtex4-2}
\usepackage[utf8]{inputenc}
\usepackage{amsmath}
\usepackage{amssymb}
\usepackage{amsfonts}
\usepackage{graphicx}
\usepackage{siunitx}
\usepackage{color}
\usepackage[colorlinks=true,citecolor=blue,urlcolor=red]{hyperref}
\usepackage{orcidlink}

\newcommand{\B}{\mathrm{B}}
\newcommand{\D}{\mathrm{D}}
\newcommand{\tip}{\mathrm{tip}}

\newcommand{\dx}{\d x \,}
\newcommand{\dy}{\d y \,}
\newcommand{\dX}{\d X \,}
\newcommand{\dY}{\d Y \,}

\renewcommand{\d}{\mathrm{d}}

\begin{document}

\title{Multimode characterization of an optical beam deflection setup}

\author{Alex Fontana\,\orcidlink{0000-0001-5522-3584}}
\author{Ludovic Bellon\,\orcidlink{0000-0002-2499-8106}}
\email{ludovic.bellon@ens-lyon.fr}
\affiliation{\href{https://ror.org/02feahw73}{CNRS}, \href{https://ror.org/04zmssz18}{ENS de Lyon}, \href{https://ror.org/00w5ay796}{Laboratoire de Physique}, F-69342 Lyon, France}
\date{\today}

\begin{abstract}
Optical beam deflection is a popular method to measure the deformation of micromechanical devices. As it measures mostly a local slope, its sensitivity depends on the location and size of the optical spot. We present a method to evaluate precisely these parameters, using the relative amplitude of the thermal noise induced vibrations. With a case study of a microcantilever, we demonstrate the accuracy of the approach, as well as its ability to fully characterize the sensitivity of the detector and the parameters (mass, stiffness) of the resonator.
\end{abstract}

\maketitle

\section{Introduction}
To measure the minute deformations of micrometer-sized mechanical systems, such as cantilevers or membranes, optical beam deflection (OBD)~\cite{Jones1961, Meyer1988} is a first choice method: it combines ease of implementation and high sensitivity. It is used for example in mass~\cite{Dohn-2005}, acoustic~\cite{Ren-2021}, chemical~\cite{Lavrik-2004} and bio~\cite{Tamayo-2013} sensing, and is the ubiquitous method to measure deflection of the cantilever probe in atomic force microscopy~\cite{Giessibl2003,Butt-2005}. As such, many articles are dedicated to describing its sensitivity, calibration, and limitations~\cite{Putman-1992, Gustafsson1994, Schaffer1998, Stark-2004, Schaffer2005, Fukuma-2005, Beaulieu-2006, Beaulieu-2007, Herfst-2014}.

Some of the key points in assessing the sensitivity of the OBD are the laser spot location and size on the micro-mechanical device. Those two parameters play an important role, especially if high order oscillation modes are targeted. Indeed, the method is sensitive to the local slope of the reflective surface, thus it will fail if the deformation implies a flat slope at the measurement point, or it will lose precision if the slope varies significantly under the spread of the spot~\cite{Schaffer2005}. Although these quantities can sometimes be precisely adjusted in the experiment, most of the time they are just not controlled and hidden parameters of a global calibration procedure.

We devise in this paper a precise way to estimate the spot location and size on a case study of a cantilever. We use for this purpose a single Thermal Noise Measurement (TNM)~\cite{Butt1995} of several resonance modes, including deformations in flexion and torsion. Since the equipartition relation prescribes the amplitude of the thermal fluctuations of these modes~\cite{Paolino-2009}, their relative magnitude is linked to the sensitivity of the OBD for each mode, from which we derive the location and size of the spot. As an extra feature, for a well-characterized optical system or using an independent measurement, one can also deduce the cantilever mass, stiffness, and the OBD sensitivity from the same single TNM.

We first present the principle of TNM for the flexural and torsional resonances of a cantilever, introducing the sensitivities that allow us to calibrate the displacements. Next, we compute an experimentally accessible quantity to deduce the position and size of the optical probe beam. We then use this method in an experiment and demonstrate how to fully characterize the OBD method (sensitivity) and the cantilever (mass, stiffness).

\begin{figure}
\begin{center}
\includegraphics[width=0.4\textwidth]{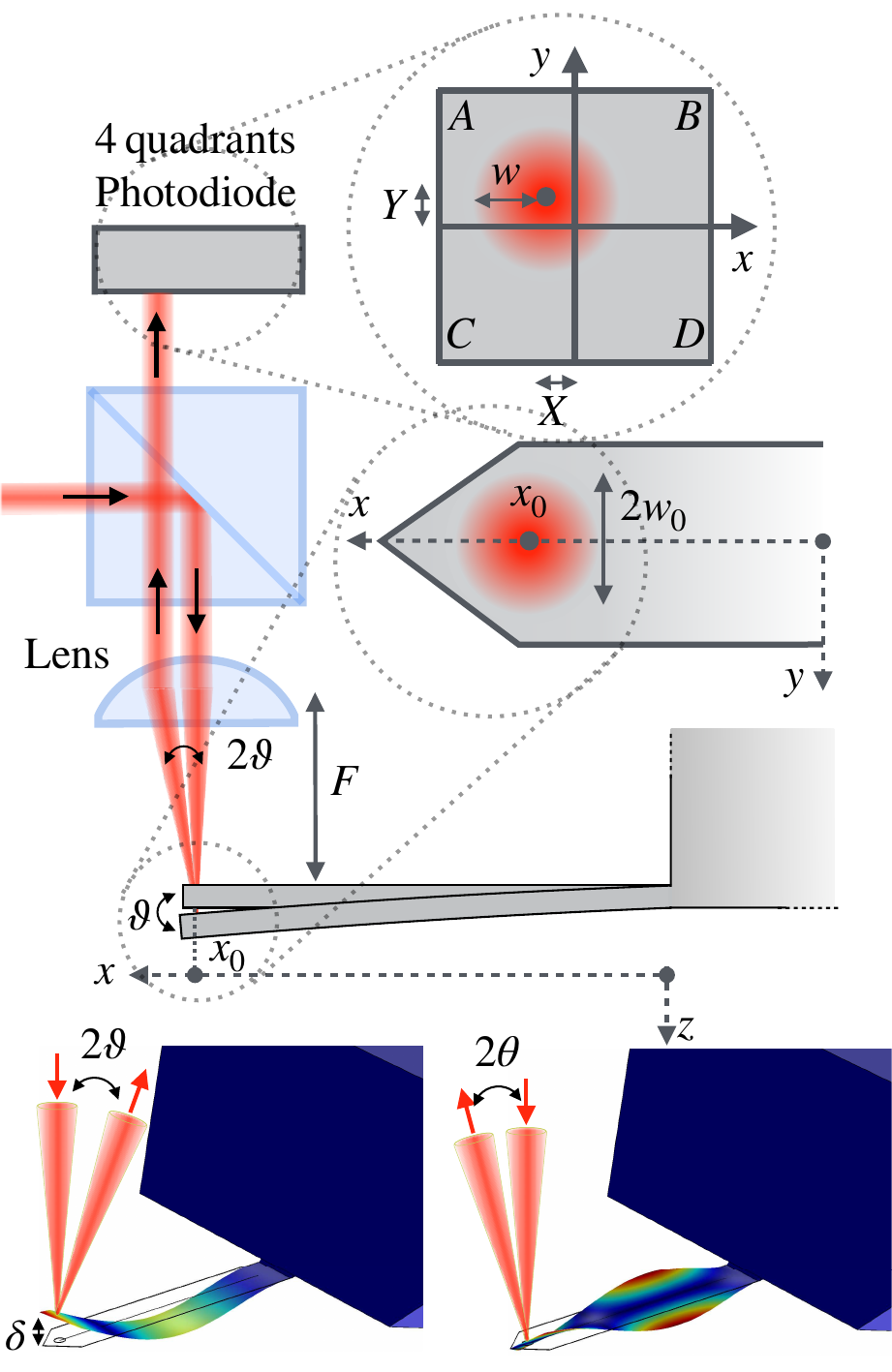}
\caption{The deformation of a micro-cantilever is measured thanks to the optical beam deflection technique: a laser beam is reflected close to the tip of the cantilever at $x_0$ with a waist size $w_0$. The beam is then redirected towards a four-quadrants photodiode in the position $X,Y$ with radius $w$. In the case of no displacement, the beam is centered on the sensor. If the cantilever bends, the beam is reflected with an angle. This angle corresponds to a shift in the $X$ direction in the photodiode in case of deflection $\delta$ and in the $Y$ direction in case of torsion $\theta$. }
\label{fig.setup}
\end{center}
\end{figure}

\section{Calibration method overview}
\subsection{OBD measurement}
The principle of measuring the deflection of a micro-cantilever with the OBD technique is depicted in Fig.~\ref{fig.setup}. A laser beam is focused on the surface of the cantilever at a position $x_0$ along its longitudinal axis, and reflected towards a 4-quadrants photodiode. In first approximation (small spot size), when the cantilever presents a deformation, the reflection occurs with an angle which is twice the local slope of the cantilever at $x_0$. We divide this angle into two components: one due to the flexion of the cantilever $\vartheta(x_0)$ (which is proportional to the vertical deflection $\delta$) and one due to its torsion $\theta(x_0)$. After passing back through the focusing lens, the reflected beam is shifted in the $x-y$ plane of the 4 quadrants photodiode:
\begin{equation}
\label{eq:XY}
X = 2 \mu F \vartheta(x_0), \qquad Y = 2 \mu F \theta(x_0),
\end{equation}
with $F$ the focal length of the lens, $X,Y$ the distance from the center of the sensor (with proper initial centering of the laser beam), and $\mu$ a magnification factor depending on the details of optical system. For example, $\mu=1$  for the optical scheme of Fig.~\ref{fig.setup} when the cantilever is at the focal point of the lens, but $\mu F$ should be replaced by the distance from the tip to the sensor if the latter collects directly the reflected beam. Each quadrant records an incoming power, namely $A, B, C, D$, from which we evaluate two contrasts:
\begin{equation}
\begin{split}
\label{eq.Cx2}
C_x &\equiv \frac{(B+D)-(A+C)}{A+B+C+D} \\
C_y &\equiv \frac{(A+B)-(C+D)}{A+B+C+D}.
\end{split}
\end{equation}
These dimensionless quantities are the raw signal available to detect the flexion and torsion of the cantilever with an OBD. In many commercial AFMs, these signals are available as voltages (typically $C_x \times \SI{10}{V}$ for deflection). For small displacements, they are proportional to the spot position $(X,Y)$ on the photodiode. Indeed, let us consider a gaussian laser beam with the following irradiance profile at the photodetector surface:
\begin{equation}
I(x,y) = I_0 e^{-2\frac{(x-X)^2+(y-Y)^2}{w^2}},
\end{equation}
with $w$ the $1/e^2$ radius of the beam. If the sensor size is much larger than the beam size, we directly get:
\begin{equation}
\begin{split}
\label{eq.Cxcomp}
C_x &= \mathrm{erf}\left(\frac{\sqrt{2} X}{w}\right) \approx \frac{2}{\sqrt{\pi}} \frac{\sqrt{2} X}{w} = 4\mu \sqrt{\frac{2}{\pi}} \frac{F}{w} \vartheta(x_0), \\
C_y &= \mathrm{erf}\left(\frac{\sqrt{2} Y}{w}\right) \approx \frac{2}{\sqrt{\pi}} \frac{\sqrt{2} Y}{w} = 4\mu \sqrt{\frac{2}{\pi}} \frac{F}{w} \theta(x_0).
\end{split}
\end{equation}
The approximation is valid for small displacements around the center of the photodiode $X_0,Y_0=0$, i.e. $X,Y\ll w$. Using diffraction laws for the Gaussian laser beam, the ratio $F/w$ is directly given by the beam waist $w_0$ at the focal point: $F/w=\pi w_0 /\lambda$, where $\lambda$ is the light wavelength. Eventually, the local angles are recovered at $x=x_0$ from the measured contrasts $C_{x,y}$.

\begin{figure}
\begin{center}
\includegraphics{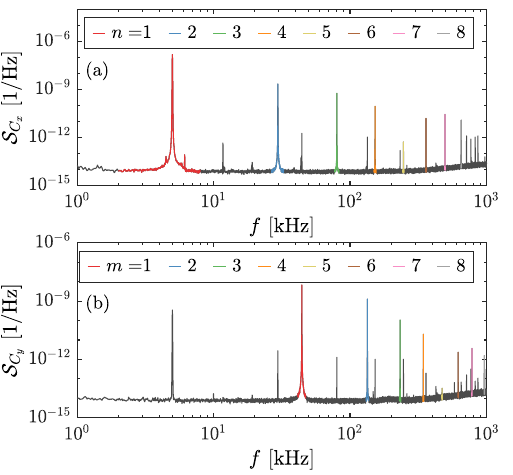}
\caption{(a) PSD of $C_x$ on an Arrow TL8~\cite{TL8} cantilever in vacuum. The thermal noise appears as resonance peaks at flexural frequencies $f_n$ with no overlap. The amplitude of each one is evaluated through an integral of the PSD in a small frequency range around each resonance. (b) PSD of $C_y$, with the torsional resonances at $f_m$ in evidence. }
\label{fig.spectra}
\end{center}
\end{figure}

\subsection{Sensitivity}
The next step is to infer the actual deflection $\delta$ or torsion $\theta$ at the end of the cantilever $x=L$, where the tip is located. We therefore need a model for the deformation profile along the cantilever. In Fig.~\ref{fig.spectra}, we plot an example of the Power Spectrum Density (PSD) of $C_x$ and $C_y$ acquired during a TNM on a cantilever Arrow TL8~\cite{TL8} in vacuum. These PSDs can be seen as a sum of resonances with no overlap: the deformation is the superposition of the eigenmodes of the mechanical beam for the considered motion. The deflection $\delta(x,t)$ (resp. torsion $\theta$) can be decomposed into the solutions $\phi_n(x)$ of the Euler-Bernoulli equation (resp. $\phi_m(x)$ of the Barr equation, see Appendix~\ref{appendix.phin} and \ref{appendix.phim} for more details):
\begin{align}
\delta(x,t) &= \sum_n \delta_n \phi_n(x) e^{i\omega_n t}, \label{eq.deltan}\\
\theta(x,t) &= \sum_m \theta_m \phi_m(x) e^{i\omega_m t},\label{eq.deltam}
\end{align}
where from now on $n$ (resp. $m$) stands for the mode number for deflection (resp. torsion), and $\omega_{n,m}$ is the resonance angular frequency. For each mode, we can thus link the slopes $\vartheta(x_0,t)=\partial_x \delta(x_0,t)$ and $\theta(x_0,t)$ at position $x_0$ to the deformation amplitude $\delta_n, \theta_m$. Therefore, we can define the sensitivities $\sigma_{n, m}(x_0,w_0)$ linking $\delta_n, \theta_m$ to the contrast amplitudes $C_{n, m}$:
\begin{equation}
\label{eq.Cequalsigmadelta}
\begin{split}
C_n = \sigma_n \delta_n, \qquad C_m = \sigma_m \theta_m,
\end{split}
\end{equation}
with
\begin{equation}
\label{eq.sigmaeasy}
\begin{split}
\sigma_n (x_0,w_0) &= \frac{4 \mu \sqrt{2\pi} w_0}{\lambda} \frac{\d\phi_n}{\d x} (x_0),\\
\sigma_m (x_0,w_0) &=  \frac{4 \mu \sqrt{2\pi} w_0}{\lambda} \phi_m (x_0).
\end{split}
\end{equation}

As shown by Eqs.~\ref{eq.sigmaeasy}, in order to maximize $\sigma_{n, m}$ one should maximise the laser spot radius $w_0$ on the cantilever~\cite{Gustafsson1994}. Nevertheless, the approximation of a Gaussian reflected beam with an angle equal to twice the local slope at $x_0$ ceases to hold in this case, especially at large mode number where the radius $w_0$ is comparable to the wavelength of the mode shape: the beam probes a nonuniform slope on the cantilever. In order to compute the flexural sensitivity $\sigma_n$  in the general case, we need to compute the diffraction of the light field $E(x-x_0,y,w_0) = E_0 e^{-[(x-x_0)^2+y^2]/w_0^2}$ reflected by the cantilever. To include the effect of the triangular tip of our sample (see Fig.~\ref{fig.setup}), we describe the geometry of the cantilever by its length $L$, uniform thickness $H$, and position-dependent width $W(x)$. We then follow Refs.~\onlinecite{Schaffer1998,Schaffer2005}, and show that: 
\begin{equation}
\label{eq.sn}
\begin{split}
\sigma_n &= \frac{4\mu}{\lambda S} \int_0^{L} \d x \int_0^{L}\d x^\prime \int_{-\frac{\mathrm{min}\left(W(x),W(x')\right)}{2}}^{\frac{\mathrm{min}\left(W(x),W(x')\right)}{2}} \d y \\
& E(x-x_0,y,w_0) E(x^\prime-x_0,y,w_0) \frac{\phi_n(x)-\phi_n(x^\prime)}{x-x^\prime},
\end{split}
\end{equation}
with
\begin{equation}
S = \int_0^{L} \d x \int_{-\frac{W(x)}{2}}^{\frac{W(x)}{2}} \d y \, E(x-x_0,y,w_0)^2
\end{equation}
the total light power collected by the photodiode. The torsion sensitivity $\sigma_m$ is similarly expressed as:
\begin{equation} \label{eq.sm}
\begin{split}
\sigma_m &= \frac{4\mu}{\lambda S} \int_0^{L} \dx \phi_m(x) \\
& \left| \!\int_{-\frac{W(x)}{2}}^{\frac{W(x)}{2}} \dy E(x-x_0,y,w_0)\right|^2.
\end{split}
\end{equation}
The calculations are shown in Appendix \ref{appendix.diffraction}.

In the limit of small spot size $w_0$, the electric field contribution is equivalent to Dirac's distribution centered in $x_0$, so that the above expressions simplify to Eqs.~\ref{eq.sigmaeasy}. These formulas have a direct analog when considering static deformation instead of resonant modes: we only need to replace the mode shape $\phi_{n,m}(x)$ by the static deformation profile $\phi_s(x)$. The static sensitivity $\sigma_s (x_0,w_0)$ is then the one commonly calibrated in an AFM with a force curve on a hard surface (invOLS, usually in $\SI{}{V/nm}$), allowing to convert the photodetector output to a static deflection value. Once the deformation profile is set, only two parameters are needed for the calibration of the sensitivity: the laser spot position $x_0$ and size $w_0$. Although these quantities can sometimes be measured in the experiment, we now discuss a method to retrieve them from the thermal noise measurement itself. 

\subsection{TNM calibration}
When the cantilever is in thermal equilibrium at a temperature $T$, the equipartition principle writes
\begin{subequations}
\label{eq.equipartition}
\begin{align}
   M \omega_n^2 \langle \delta_n^2\rangle = k_n \langle \delta_n^2 \rangle = k_\B T.\\
   J \omega^2_m \langle \theta_m^2 \rangle  = c_m \langle \theta_m^2 \rangle = k_\B T.
\end{align}
\end{subequations}
with $M$ the mass and $J$ the moment of inertia of the cantilever, $k_n$ the stiffness of mode $n$, $c_m$ the torsional stiffness of mode $m$ and $k_\B$ the Boltzmann constant (see Appendix \ref{sec.equipartition} for the derivation). The stiffness $k_n$ can be expressed with the spatial eigenvalue $\alpha_n$ of mode $n$:
\begin{equation} \label{eq.defkn}
M\omega_n^2 \equiv k_n = \gamma K \alpha_n^4,
\end{equation}
with $K$ the static stiffness of the cantilever, $k_n$ the mode stiffness and $\gamma\sim 1/3$ a geometrical constant (see Appendix \ref{appendix.phin}). A similar equation can be derived for the torsional stiffness $c_m$. The quantities $\langle \delta_n^2\rangle, \langle \theta_m^2\rangle$ represent the amplitude of fluctuations for the two deformations, i.e. the thermal content of each mode. From Eqs.~\ref{eq.Cequalsigmadelta}, we can then write that the amplitude of the measured contrasts is:
\begin{equation}
\label{eq.thmeas}
\begin{split}
\langle C_n^2 \rangle & = \frac{\sigma_n^2(x_0,w_0)}{M \omega_n^2} k_\B T = \frac{\sigma_n^2(x_0,w_0)}{k_n} k_\B T, \\
\langle C_m^2 \rangle & = \frac{\sigma_m^2(x_0,w_0)}{J \omega_m^2} k_\B T= \frac{\sigma_m^2(x_0,w_0)}{c_m} k_\B T.
\end{split}
\end{equation}
Experimentally, the angular frequencies $\omega_{n, m}=2\pi f_{n, m}$ are easily extracted from a Lorentzian fit of the resonance in the thermal noise spectrum (see fig.~\ref{fig.spectra}). $\langle C_n^2 \rangle$ (resp. $\langle C_m^2 \rangle$) is measured by integrating the PSD $\mathcal{S}_{C_x}$ (resp. $\mathcal{S}_{C_y}$) on an adequate frequency range $2 \Delta f$ around $f_{n}$ (resp. $f_m$):
\begin{equation}
\label{eq.Cn}
\langle C_{n, m}^2 \rangle = \int_{f_{n, m}-\Delta f}^{f_{n, m}+\Delta f} \d f\,\mathcal{S}_{C_{x, y}}(f).
\end{equation}
During the choice of $\Delta f$ and the integration one should take care to remove the contribution of the flat background noise, which is however negligible in most cases. From repeated measurements of the thermal noise spectra, one can also extract the statistical uncertainty $\Delta \langle C_{n, m}^2 \rangle$ of the amplitude of each mode. In our experiment, this quantity decreases from $\Delta \langle C_{2}^2 \rangle / \langle C_{2}^2 \rangle = \SI{23}{\%}$ for mode $n=2$ to roughly $\SI{5}{\%}$ for higher modes for 100 spectra evaluated of $\SI{2}{s}$ long datasets (see Appendix \ref{appendix.errmodes}).

\begin{figure*}[t]
\begin{center}
\includegraphics{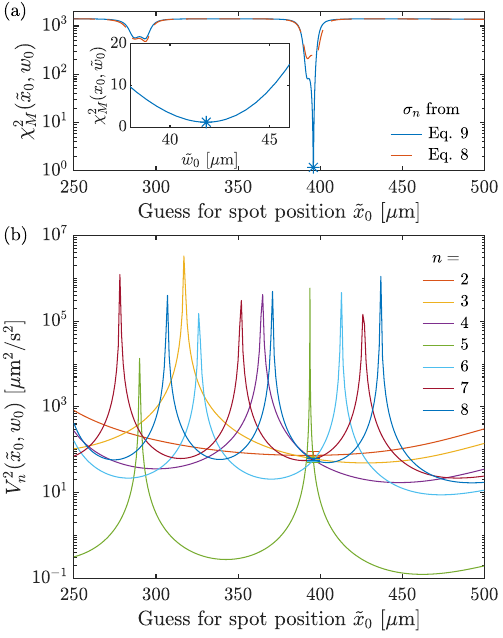} \hfill \includegraphics{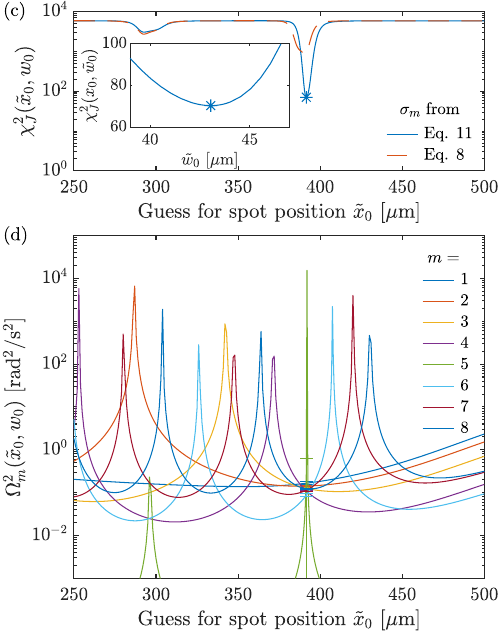}
\caption{Estimation of $x_0$ and $w_0$. (a) For the flexion modes, $\chi_M^2$ presents a global minimum in $x_0 = (395.9\pm 0.2)\SI{}{\mu m}$ (main figure) and $w_0=(41.8\pm 1.4)\SI{}{\mu m}$ (inset). If the sensitivity $\sigma_n$ is evaluated with the local slope model (Eq.~\ref{eq.sigmaeasy}, dashed line) instead of the diffraction integral (Eq.~\ref{eq.sn}, plain line), the minimum of $\chi_M^2$ is approximately is at the same position, but far less sharp: $(393\pm 5)\SI{}{\mu m}$. (b) The velocity variances $V_n^2(\tilde x_0,w_0)$, plotted here as a function of $\tilde x_0$ for $w_0$ fixed, intersect within uncertainties only at the location of the laser spot. (c) For torsion modes, $\chi_J^2$ present a global minimum in $x_0 = (392.0 \pm 0.3)\SI{}{\mu m}$ (main figure) and $w_0=(43.0 \pm 0.9)\SI{}{\mu m}$ (inset). The sharp minimum is blurred with the local slope model (Eq.~\ref{eq.sigmaeasy}, dashed line) with respect to the full diffraction model for $\sigma_m$ (Eq.~\ref{eq.sm}, plain line). (d) The angular velocity variances $\Omega^2_m(\tilde x_0,w_0)$, plotted here as a function of $\tilde x_0$ for $w_0$ fixed, intersect within uncertainties only at the location of the laser spot.}
\label{fig.err}
\end{center}
\end{figure*}

Let us next define the quantities $V_{n}^2$, $X_n^2$ and $\Omega_{m}^2$ by:
\begin{subequations}
\begin{align}
V_n^2(\tilde x_0, \tilde w_0) & \equiv \frac{\omega_n^2 \langle C_n^2 \rangle}{\sigma_n^2(\tilde x_0,\tilde w_0)} =  \frac{k_B T}{M}, \label{eq.Vn2}\\
X_n^2(\tilde x_0, \tilde w_0) & \equiv \gamma \frac{\alpha_n^4 \langle C_n^2 \rangle}{\sigma_n^2(\tilde x_0,\tilde w_0)}=  \frac{k_B T}{K}, \label{eq.Xn2}\\
\Omega_m^2(\tilde x_0, \tilde w_0) & \equiv \frac{\omega_m^2 \langle C_m^2 \rangle}{\sigma_m^2(\tilde x_0,\tilde w_0)} =  \frac{k_B T}{J}, \label{eq.Om2}
\end{align}
\end{subequations}
where the last equalities hold when the guesses $\tilde x_0, \tilde w_0$ of the position and beam radius matches the experimental values $x_0,w_0$. For those values then, $V_{n}^2(x_0,w_0)$ ($X_n^2$, $\Omega_m^2$) is independent of the mode number $n$ or $m$ and is a velocity (resp. deflection, angular velocity) variance common to all modes. If the calibration is performed in a fluid, the cantilever drags some material while vibrating~\cite{Sader1998}, thus a mode-dependent mass should be considered. In this case, it is recommended to work with Eq.~\ref{eq.Xn2} as the static stiffness $K$ is mode-independent. In our experiment, performed in vacuum, the mass of the cantilever $M$ is mode independent, and the use of Eq.~\ref{eq.Vn2} is equally suitable. In the following, we only consider $V_n^2$ and $\Omega_m^2$.

Using the TNM, we measure the numerator of Eqs.~\ref{eq.Vn2} and \ref{eq.Om2}, and using Eqs.~\ref{eq.sn} and \ref{eq.sm}, we compute the denominator as a function of $\tilde x_0, \tilde w_0$. When plotting $V_{n}^2$ and $\Omega_m^2$, all modes cross at $x_0,w_0$, as illustrated in Fig.~\ref{fig.err}.

To estimate the values of $x_0$, $w_0$, and $M$ from a fitting procedure, we can minimize the following $\chi_M^2$ function:
\begin{equation}
\label{eq.chi2}
\chi^2_M(\tilde x_0, \tilde w_0, M) \equiv \sum_n \left(\frac{V_{n}^2 (\tilde x_0, \tilde w_0)-k_B T/M}{\Delta V_{n}^2(\tilde x_0, \tilde w_0)}\right)^2,
\end{equation}
where $\Delta V_{n}^2$ is the uncertainty on $V_{n}^2$, which is mainly due to the propagation of the statistical uncertainty of $\langle C_n^2 \rangle$ through Eq.~\ref{eq.Vn2} (the uncertainty on the frequency is very small). An equivalent $\chi^2_J$ function can obviously be defined for torsion, replacing $n,M$ by $m,J$. Since Eq.~\ref{eq.chi2} is quadratic in $1/M$, minimization in $M$ can be done analytically, leading to:
\begin{equation} \label{eq.M}
M=k_B T \frac{\sum_n (\Delta V_{n}^2(\tilde x_0, \tilde w_0))^{-2}}{\sum_n V_{n}^2(\tilde x_0, \tilde w_0)(\Delta V_{n}^2(\tilde x_0, \tilde w_0))^{-2}},
\end{equation}
which can be recasted in Eq.~\ref{eq.chi2} to remove the dependency of $\chi^2_M$ in $M$. The minimisation of $\chi^2_M$ then leads to the most probable value of $\tilde x_0 = x_0$ and $\tilde w_0 = w_0$. Since $k_B T/M$ is calculated as a weighted average of $V_n^2$, with weights $(\Delta V_n^2)^{-2}$, the statistical uncertainty of the mass can be easily computed from:
\begin{equation} \label{eq.vM}
\mathrm{Var}\left(\frac{k_B T}{M}\right) = \frac{1}{\sum_n (\Delta V_{n}^2(\tilde x_0, \tilde w_0))^{-2}}.
\end{equation}
In our experiment, we get a relative uncertainty of $\SI{2.7}{\%}$.

Once the laser spot position and size are known thanks to this contactless TNM, we can in principle easily compute all parameters of the cantilever: the mass $M$ with Eq.~\ref{eq.M}, the stiffness $k_n$ of each mode with Eq.~\ref{eq.defkn}, from which we deduce the static stiffness $K$. We can also estimate the photodetector sensitivity for one vibration mode or for a static deformation with Eq.~\ref{eq.sn}. These last steps require the knowledge of the optical magnification $\mu$ of the setup, which can be difficult to assess due to imprecise tuning or non-idealities (cantilever not at the focal point of the laser beam, aberrations or aperture diffraction in the optical system). In contrast, if one of these parameters ($M$, $K$) is known from another method, it can be used to infer $\mu$ and the sensitivities of the OBD. This includes, for example, the use of the Sader model~\cite{Sader1998} in atomic force microscopy to evaluate the stiffness of the first oscillation mode $k_1$. Another example is given in the following on our specific sample operated in vacuum (thus not compatible with the Sader method), using the resonance frequencies to determine the mass $M$~\cite{Lubbe-2012}.

\section{Experiment}
This calibration procedure has been commonly used in previous works in our group in order to extract a calibrated measurement of the thermal noise of the cantilever~\cite{Fontana-2020}. We illustrate it here with an Arrow TL8~\cite{TL8} silicon cantilever, $L=\SI{500}{\mu m}$ long, $W=\SI{95}{\mu m}$ wide and $H\sim\SI{1}{\mu m}$ thick, with a triangular free end. In the experiment, we focus the probe near the tip of the cantilever with a diameter close to $W$, as suggested in Fig.~\ref{fig.setup}. 

We report in Fig.~\ref{fig.err} the measurement of $x_0$ and $w_0$ through the minimisation of $\chi^2$, deduced from a single TNM. As we can see, $\chi^2$ has a well defined minimum at a certain position $x_0$ and radius $w_0$ where the velocity variance $V_{n}$ or $\Omega_m$ converge for all the measured modes. For the fitted laser position $x_0$, we note that there is only a $\SI{1}{\%}$ difference between torsion and flexion. The estimated beam size $w_0$ are also very similar for the 2 mode families ($\SI{3}{\%}$ difference, within uncertainties).

\begin{figure}
\begin{center}
\includegraphics{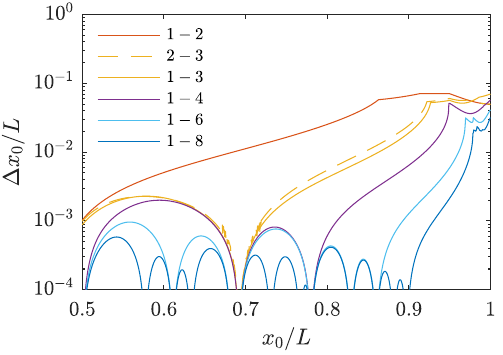} 
\caption{Uncertainy on the estimated position $\Delta x_0$ as a function of the measurement position $x_0$. The curves demonstrate the effect of the number $N$ of modes available (from 2 to 8 from top to bottom): the larger $N$, the better is the precision. It can be noted that higher order modes are the most useful, and that mode 1 has little influence on the result: using mode 2 and 3 only (dashed yellow) leads to a result very similar to using mode 1 to 3 (plain yellow). The proximity of a sensitivity node is highly beneficial and lowers significantly the uncertainty. In this figure, $\Delta x_0$ is computed by propagating a $\SI{5}{\%}$ uncertainty on all $\langle C_n^2 \rangle$ on a model case corresponding to a rectangular cantilever in the limit of small detection spot size.}
\label{fig.modenumberstudy}
\end{center}
\end{figure}

We exclude from the procedure the first mode in flexion ($n=1$), whose thermal noise amplitude is corrupted by spurious phenomena (mainly self oscillations of the cantilever due to an opto-mechanical coupling~\cite{Fontana-2020,Fontana-2021}). Albeit being the most prominent, mode $n=1$ is the least sensitive to small changes in the probing position or waist, thus its exclusion in our experiment is not problematic. Conversely, at high pressure, or with different probes, self-oscillations should not take place, thus the first flexural resonance will most likely be available. We report the effect of limiting the number of resonances considered in Fig.~\ref{fig.modenumberstudy}.

To compute the sensitivity of our OBD, we need the value of $\mu$, which is not well controlled in our setup, as the cantilever is not exactly in the focal plane of the lens. From  the geometry of our setup and Gaussian beam optics using the value of $w_0$ measured above, we estimate $\mu\sim0.5$. Instead, we use an independent measurement of the cantilever mass $M$ to assess $\mu$. This estimation of $M$ is inspired by Ref.~\onlinecite{Lubbe-2012}: the angular resonance frequencies $\omega_n$ are linked to the spatial eigenvalues $\alpha_n$ of the Euler-Bernoulli equations by the dispersion relation
\begin{equation} \label{eq.disprelEB}
    \omega_n=\sqrt{\frac{Y}{12\rho}}\frac{H}{L^2}\alpha_n^2
\end{equation}
where $Y=\SI{169}{GPa}$ and $\rho=\SI{2330}{kg/m^3}$ are the Young modulus and density of silicon (the material of the cantilever). Since the eigenvalues $\alpha_n$ are known from computing the resonant mode $\phi_n$, $L$ is measured with an optical microscope, and the $\omega_n$ are measured in the TNM, $H$ can be evaluated independently for all modes, yielding $H=(0.707\pm0.004)\SI{}{\mu m}$. Interestingly, the dispersion relation of the torsion modes~\cite{Barr1962} can also be used to estimate $H=(0.687\pm0.007)\SI{}{\mu m}$, in very good agreement with flexion. Knowing the thickness, the density and the plane geometry ($L$, $W$, triangular tip) lead to the mass $M=\SI{6.9e-11}{kg}$ with a good precision. Finally, we can adjust the magnification to $\mu=0.6$ for Eq.~\ref{eq.Vn2} to match this value.

Now that the geometry of the cantilever is known, other quantities like the static stiffness $K=\SI{0.011}{N/m}$, or the dynamic ones $k_n$ can be computed. The moment of inertia of the cantilever can also be evaluated to $J=\SI{4.6e-20}{m^2.kg}$. It falls reasonably close to the value extracted from Eq.~\ref{eq.Om2}, $J=\SI{3.4e-20}{m^2.kg}$. This slight underestimation of $J$ by the torsion modes is attributed to the difficulty in accurately describing the shape of the torsional mode $\phi_m$ close to the triangular tip.

\section{Conclusions}
In our experiment, we take advantage of the many resonant modes that are available to achieve this calibration. In order to show the feasibility of such a procedure in other experimental setups, let us discuss briefly two less ideal cases. First, if the beam size is small compared to the wavelength of the highest normal mode $N$ considered, i.e. $2w_0<L/N$ (see Fig.~\ref{fig.phinm}), its estimation becomes difficult. However, in such case $w_0$ has little influence on the sensitivity dependence of the modes, and can be fixed to an approximate value without influencing the accuracy on the measurement of $x_0$. Using an independent measurement of the mass then yields the sensitivity and thus the optical beam size. Second, if fewer resonances are available, the estimation of the probe's position becomes less precise (see Fig. \ref{fig.modenumberstudy}). This is the case of experiments performed in fluids (decreasing the quality factor leads to worse signal-to-noise ratios) or with different kinds of cantilevers (shorter ones shift the resonances at high frequency). If $N$ is too small, the error on the estimation of the probing point (and thus on the stiffness) will be high. It can be emphasized that if the precision required on $x_0$ is high, then it is advisable to choose a location close to a node in sensitivity, as we did is our case study using mode 5. In general, the precision drops when $x_0$ get close to $L$, and is larger than all available nodes.

When $N$ is sufficiently large, we end up with a full characterization of our OBD sensor. Indeed, the location and size of the laser spot on the micro-mechanical resonator, the sensitivity of the measurement, the geometry, mass and stiffness of the cantilever are all assessed. Let us then recapitulate the calibration steps:
\begin{enumerate}
\item Measure the thermal noise amplitude of all the available, say flexural, resonances $\langle C_n^2 \rangle$ and the associated resonance frequencies $\omega_n$.
\item Compute the sensitivity $\sigma_n$ assuming a magnification $\mu=1$.
\item Compute the velocity variance $V_n^2$ for a measurement in vacuum, or the displacement variance $X_n^2$ for a measurement in fluid.
\item Minimize Eq.~\ref{eq.chi2} in order to obtain the probing spot $x_0$ and beam size $w_0$.
\item Extract the mass $M$ (or the stiffness $K$) from an independent measurement.
\item Finally, deduce the actual magnification $\mu$ from Eq.~\ref{eq.Vn2} and obtain the adjusted calibration function $\sigma_n$. 
\end{enumerate}
Following this procedure, the measurement device is fully calibrated.

As a final conclusion, let us mention that we focused in this article on the OBD configuration, but the calibration of other optical detection schemes such as interferometry could also be assessed in the same way. In this case, the sensitivity nodes are the zeros of the eigenmodes and are closer to the tip, further expanding the method precision in detecting the probing point. Furthermore, the intrinsic calibration of the interferometer would alleviate the need of an independent measurement of one oscillator property.

\section*{Acknowledgments}
This work has been financially supported by the French r\'egion Auvergne Rh\^one Alpes through project Dylofipo and the Optolyse plateform (CPER2016), and by the Agence Nationale de la Recherche through grants ANR-18-CE30-0013 and ANR-22-CE42-0022. We thank Nicolas Barros for useful complementary data from a finite element method simulation to validate our analytical model.

\section*{Data availability}
The data that supports the findings of this study and the scripts to compute mode shapes and perform all calibration steps are openly available in Ref.~\onlinecite{Fontana-2024-Dataset}.

\newpage
\appendix

\section*{Appendix}

\section{Deflection normal modes}\label{appendix.phin}
\begin{figure}[b]
\begin{center}
\includegraphics{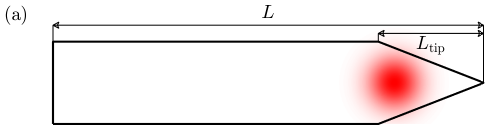} 
\includegraphics{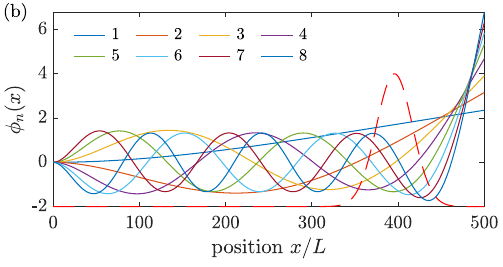} 
\includegraphics{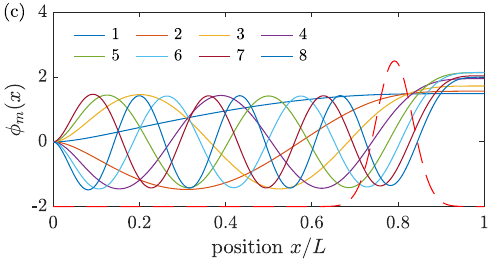} 
\caption{Eigenmodes computed for the triangular tipped cantilever. (a) Sketch of the cantilever (top view), with a Gaussian beam at $x_0=\SI{396}{\mu m}$ of $1/e^2$ radius $w_0=\SI{42}{\mu m}$. The profile of this spot is plotted in dashed red line on the bottom graphs. It illustrates the non constant slope of the high order normal modes in the light beam. (b) $\phi_n(x)$ are the solutions of the Euler-Bernoulli Eq.~\ref{eq.EB} for $n=1$ to $8$ with a non uniform width characterized by $l_\tip=L_\tip/L=0.245$. The boundary conditions are clampled-free, and solved by numerical integration. With respect to a rectangular cantilever, the deflection of the free end is amplified by the weakening local rigidity when $W$ decreases. (c) $\phi_m(x)$ are the solutions of Barr's Eq.~\ref{eq.EB} for $n=1$ to $8$. Note that $\phi_m$ represent here directly the slope of the cantilever in the transverse direction, while in panel (b) $\phi_n$ is the vertical deflection, the slope of which is sensed by the OBD.}
\label{fig.phinm}
\end{center}
\end{figure}

The Euler-Bernoulli equation~\cite{Landau1970} for the normal modes of the cantilever vertical deflection $\delta$ can be written as:
\begin{equation}
\label{eq.EB}
\left[\rho WH \frac{\partial^2}{\partial t^2} + \frac{\partial^2}{\partial x^2 }Y \frac{WH^3}{12} \frac{\partial^2}{\partial x^2}\right]\delta(x,t) = 0,
\end{equation}
with $\rho$ the density of the material, $Y$ the Young modulus, and $W$ and $H$ the width and thickness of the cantilever. The following boundary conditions apply (clamped-free beam): $\delta(0,t)=0$, $\partial_x\delta(0,t)=0$, $\partial_x^2\delta(L,t)=0$, and $\partial_x WH^3\partial_x^2\delta(L,t)=0$.
The triangular tip of the cantilever is taken into account by the $x$-dependent width $W(x)$:
\begin{equation}
W(x) = \min\left(\frac{L-x}{L_\tip}W,W\right), 
\end{equation}
with $L=\SI{500}{\mu m}$ and $L_\tip=\SI{82}{\mu m}$ for the sample of this case study. This boundary value problem has no simple analytical solution but can be integrated numerically, yielding the eigenmodes $\phi_n(x)$ [Fig.~\ref{fig.phinm}(b)], and the associated spatial eigenvalue $\alpha_n$ (Tab.~\ref{tab.disprelEB}). 

\begin{table}[b]
    \centering
    \begin{tabular}{|c|c|c|c|}\hline
       Mode & Eigenvalue & Freq. [kHz] & Thickness [nm]\\
       $n$ & $\alpha_n$ & $\omega_n/(2\pi)$ & $H=L^2 \sqrt{12\rho/Y} \omega_n/\alpha_n^2$\\ \hline
       1 & 2.118 & 4.992 & 711 \\\hline
       2 & 5.192 & 29.59 & 701 \\\hline
       3 & 8.529 & 80.09 & 703 \\\hline
       4 & 11.74 & 152.6 & 708 \\\hline
       5 & 14.89 & 246.2 & 710 \\\hline
       6 & 18.03 & 360.5 & 709 \\\hline
       7 & 21.18 & 496.4 & 707 \\\hline
       8 & 24.33 & 652.6 & 705 \\\hline
    \end{tabular}
    \caption{Height first computed eigenvalues of a triangular tipped cantilever with $L_\tip/L=0.245$, measured resonance frequency, and evaluation of the thickness $H$ of our sample from the dispersion relation Eq.~\ref{eq.disprelEBappendix}. All modes yield the same results, within a $\SI{0.3}{\%}$ standard deviation.}
    \label{tab.disprelEB}
\end{table}

In the straight part of the cantilever, we derive the dispersion relation between $\alpha_n$ and the angular resonance frequency $\omega_n$:  
\begin{equation}
\label{eq.disprelEBappendix}
\rho \omega_n^2 = \frac{YH^2}{12} \frac{\alpha_n^4}{L^4}
\end{equation}
The ratio of the resonance frequency and the square of the eigenvalue is thus mode independent, as advertised in Eqs.~\ref{eq.defkn} and \ref{eq.disprelEB}. We check this prediction in Tab.~\ref{tab.disprelEB}: the agreement is excellent, within $\SI{0.3}{\%}$. The dispersion actually depends on the ratio $l_\tip=L_\tip/L$, which is finely tuned to minimize the standard deviation on the values of $H$. A rectangular cantilever model for example ($l_\tip=0$) lead to an overestimation of $H$ by $\SI{12}{\%}$ and a dispersion of between modes of $\SI{7}{\%}$. The value $l_\tip=0.245$ we extract from this procedure is a bit larger than expected from geometry (0.17). We believe this small deviation is due to other non-idealities of the cantilever, such as a non-uniform thickness for example.

The geometrical factor $\gamma$ in Eq.~\ref{eq.defkn} can be computed using the expression of the mass and static stiffness of the triangular tipped cantilever:
\begin{align}
    M& =\rho WHL (1-l_\tip/2)\\
    K&= \frac{Y W H^3}{4 L^3}\frac{1}{1 +l_\tip^3/2}.
\end{align}
Using the dispersion relation (Eq.~\ref{eq.disprelEBappendix}) and the definition of $k_n$ (Eq.~\ref{eq.defkn}) leads to
\begin{equation}
\gamma=\frac{M\omega_n^2}{K\alpha_n^4}=\frac{1}{3}\left(1-\frac{1}{2}l_\tip\right)\left(1+\frac{1}{2}l_\tip^3\right).
\end{equation}

\section{Torsion normal modes}\label{appendix.phim}

To describe torsion, we use the model from Barr~\cite{Barr1962}. The vertical deflection of the normal modes $\delta(x,y,t) = \theta(x) y e^{i\omega t}$ is governed by the following coupled differential equations:
\begin{align}
    \frac{\partial^2\theta}{\partial x^2} &= -\frac{\rho}{S} \omega^2 \theta + \kappa(x) \frac{\partial \psi}{\partial x},\\
    \frac{\partial^2\psi}{\partial x^2} &= -\frac{\rho}{Y} \omega^2 \psi + \frac{\kappa(x)}{\eta(x)} \left(\psi-\frac{\partial \theta}{\partial x}\right),
\end{align}
where $S=\SI{51}{GPa}$ is the shear modulus of Silicon and $\kappa$ and $\eta$ are functions of $H$ and $s(x)=W(x)/H$:
\begin{align}
    \kappa &= 1-\frac{4}{1+s^2}\left[1-6\sum_{j=0}^{\infty}\frac{\tanh \kappa_j s}{\kappa_j^5 s}\right],\\
    \eta &= \frac{H^2}{1+s^2}\left[\frac{s^2}{12}+\sum_{j=0}^{\infty}\frac{9}{\kappa_j^6}\left(\frac{\tanh\kappa_j s}{\kappa_j s}+\frac{\tanh^2 \kappa_j s}{3} - 1\right)\!\right]\!,
\end{align}
with $\kappa_j=(j+\frac{1}{2})\pi$ (and dropping the $x$ dependency to alleviate the notations). The following boundary conditions apply (clamped-free beam): $\theta(0)=0$, $\psi(0)=0$, $\partial_x\psi(L)=0$, and $\partial_x \theta(L)=\kappa(L) \psi(L)$. This boundary value problem has no simple analytical solution but can be integrated numerically, yielding the eigenmodes $\phi_m(x)$ plotted in Fig.~\ref{fig.phinm}(c). As for flexion, using the numerical integration and the experimental values of the resonance frequencies, we can estimate the thickness of the cantilever: the value agree very well with the previous estimation, with a $\SI{1}{\%}$ dispersion (Tab.~\ref{tab.disprelBarr}). 

\begin{table}[b]
    \centering
    \begin{tabular}{|c|c|c|}\hline
       Mode & Resonance freq. [kHz] & Thickness [nm]\\
       $m$ & $\omega_m/(2\pi)$ & $H$\\ \hline
       1 & 44.31 & 703 \\\hline
       2 & 134.0 & 688 \\\hline
       3 & 232.7 & 683 \\\hline
       4 & 344.5 & 682 \\\hline
       5 & 472.0 & 684 \\\hline
       6 & 619.5 & 689 \\\hline
       7 & 782.7 & 686 \\\hline
       8 & 964.5 & 681 \\\hline
    \end{tabular}
    \caption{Height first measured torsion resonance frequencies, and evaluation of the thickness $H$ of our sample from the Barr model. All modes yield the same results, within a $\SI{1}{\%}$ standard deviation.}
    \label{tab.disprelBarr}
\end{table}

\section{Relative error of the displacement}\label{appendix.errmodes}
Our raw data consists in $\sim 100$ $2s$ time traces of the contrast signals $C_x$ and $C_y$. For each time trace, we compute the power spectrum density and extract the amplitudes $\sim 100$ $\{C_{n,m}\}$ corresponding to the resonant modes. The mean quadratic values $\langle C_{n, m}^2 \rangle$ and their statistical uncertainties $\Delta \langle C_{n, m}^2 \rangle$ are evaluated from these $\sim 100$ independent estimations.

In each time trace, the amplitude of a single mode varies slowly in time, with a typical correlation time $2 Q_n / \omega_{n,m}$, which decreases with $(n,m)$. The low-frequency modes are thus probed at a lower frequency than their high-frequency counterparts, leading to a higher statistical uncertainty on their variance. Conversely, the amplitude of the thermal noise decreases with the mode number while the floor noise slightly increases. The signal-to-noise ratio thus decreases with $(n,m)$ (and if the probing point of the displacement is close to a node), thus increasing the statistical uncertainties. Overall, the relative uncertainty of all modes is similar, around $\SI{5}{\%}$.

\section{Equipartition} \label{sec.equipartition}

When the cantilever is in thermal equilibrium at a temperature $T$, the equipartition principle imply that the mean kinetic energy of each mode is $\frac{1}{2}k_B T$, with $k_B$ the  Boltzmann constant. We can therefore write
\begin{align}
\int_0^L \d x \, \frac{1}{2} \rho H W(x) \omega_n^2 \langle \delta_n^2\rangle \phi_n^2(x) &= \frac{1}{2} k_B T, \\
\int_0^L \d x \,\frac{1}{2} \rho H \frac{W^3(x)}{12} \omega_m^2 \langle \theta_m^2\rangle \phi_m^2(x) &= \frac{1}{2} k_B T.
\end{align}
We use a convenient normalisation of the normal modes such that:
\begin{align}
\int_0^L\d x \, \rho H W(x) \phi_n^2(x)  &= \int_0^L\d x \, \rho H W(x) = M, \\
\int_0^L\d x \, \rho H \frac{W^3(x)}{12} \phi_m^2(x)  &= \int_0^L\d x \, \rho H \frac{W^3(x)}{12}  = J.
\end{align}
Note that this is the common normalization of the normal modes for a rectangular cantilever where $W$ is uniform. Combing the last four equations leads to the equipartition expression of Eq.~\ref{eq.equipartition}.

\newpage
\section{Diffraction integral}\label{appendix.diffraction}

With the mode shapes $\phi_{n,m}$ computed in the previous sections, we can proceed calculating the sensitivity of the measurement in the case of a large spot, expressed by Eqs.~\ref{eq.Cequalsigmadelta}. Our working hypothesis is a Gaussian light field $E$ of $1/e^2$ radius $w_0$ incident on the cantilever:
\begin{equation}
E(x-x0,y,w_0) = E_0 e^{-\frac{(x-x0)^2+y^2}{w_0^2}},
\end{equation}
The beam reflected from the cantilever back to the sensor from point $x,y$ travels an additional distance $2 \delta(x,y)$, twice is the vertical displacement of the cantilever. If $\delta$ is small, we can express the electric field $E_\mathrm{pd}$ on the photodiode through the diffraction integral ~\cite{Schaffer2005}:
\begin{equation}
\begin{split}
& E_\mathrm{pd} (X,Y,x_0,w_0) = \frac{k}{2\pi F}\int_0^{L} \dx \int_{-\frac{W(x)}{2}}^{\frac{W(x)}{2}} \dy  \\
& \times E(x-x_0,y,w_0) e^{2ik\delta(x,y)}e^{-ikx\frac{X}{F}}e^{-iky\frac{Y}{F}},
\end{split}
\end{equation}
with $k=2\pi/\lambda$, $X,Y$ the coordinates on the sensor and $F$ its distance to the cantilever (the lens only collects this light field and propagate it unchanged to the sensor if the cantilever is in its focal plane). Referring to Fig.~\ref{fig.setup}, let us consider quadrant $A$ as an example. The collected power is:
\begin{equation}
\label{eq.AA}
A = \int_{-\infty}^0 \d X \int_0^\infty \d Y \, \left| E_\mathrm{pd}(X,Y,x_0,w_0)\right|^2,
\end{equation}
where we assumed that the photodiode is much large than the spread of the laser spot on its surface. For the deflection case, the measured contrast $C_x$ (defined in Eq.~\ref{eq.Cx2}) is calculated as the difference between the intensity collected in the left and right quadrants (henceforth referred to as $D_x$), normalised by the total intensity $S$. For the difference we can thus write:
\begin{equation}
\label{eq.Dx1}
\begin{split}
& D_x  =   \int_0^{L} \d x \int_0^{L}\d x^\prime \int_{-\frac{W(x)}{2}}^{\frac{W(x)}{2}} \d y \int_{-\frac{W(x^\prime)}{2}}^{\frac{W(x^\prime)}{2}} \d y^\prime\\
& \times  E(x-x_0,y,w_0)  E(x^\prime-x_0,y^\prime,w_0) \\
& \times e^{2ik(\delta(x,y)-\delta(x^\prime,y^\prime))} \frac{k}{2\pi F} \int_{-\infty}^{+\infty} \d Y \, e^{-ikY(y-y^\prime)/F} \\
& \times \frac{k}{2\pi F} \left(\int_0^{+\infty} \!\!-\! \int_{-\infty}^0 \right) \d X\, e^{-ikX(x-x^\prime)/F}.
\end{split}
\end{equation}
The integral over $X$ yields a Cauchy principal part (PP) and the one over $Y$ a Dirac's delta $\delta^\D$:
\begin{equation}
\begin{split}
& \frac{k}{F}\left(\int_0^{+\infty} \!\!-\! \int_{-\infty}^0 \right) \dX e^{-ikX(x-x^\prime)/F} = \mathrm{PP} \frac{2 i }{x-x^\prime}, \\
& \int_{-\infty}^{+\infty} \dY e^{-ikY(y-y^\prime)/F} = - \delta^\D(y-y^\prime).
\end{split}
\end{equation}
Furthermore, for small displacements we can write:
\begin{equation}
e^{2ik(\delta(x,y)-\delta(x^\prime,y^\prime))} \approx 1 + 2ik(\delta(x,y)-\delta(x^\prime,y^\prime)).
\end{equation}
In this case, $D_x$ reads:
\begin{equation}
\label{eq.Dn2}
\begin{split}
& D_x  =  \frac{1}{i \pi} \mathrm{PP} \int_0^{L} \d x \int_0^{L}\d x^\prime \int_{-\frac{W(x)}{2}}^{\frac{W(x)}{2}} \d y \int_{-\frac{W(x^\prime)}{2}}^{\frac{W(x^\prime)}{2}} \d y^\prime\\
& \times  E_c(x-x_0,y,w_0)  E_c(x^\prime-x_0,y^\prime,w_0) \\
& \times \frac{1+2ik(\delta(x,y)-\delta(x^\prime,y^\prime))}{x-x^\prime}\delta^\D(y-y^\prime).
\end{split}
\end{equation}
The $0^\mathrm{th}$ order of the displacement contribution is zero because the integrand is antisymmetric with respect to $x$ and $x^\prime$. Furthermore, the principal part of the integral is dropped, as the integrand is not singular in $x=x^\prime$. We consider the contribution of the flexion mode $n$ $\delta(x,y) = \delta_n \phi_n(x)$ (from Eq.~\ref{eq.deltan}) which is independent on $y$. We then insert $\delta$ into Eq.~\ref{eq.Dn2} and integrate the Dirac's delta, now $D_x$ is renamed $D_n$ (specific to mode $n$):
\begin{equation}
\label{eq.Dn}
\begin{split}
& D_n  =  \frac{4\delta_n}{\lambda} \int_0^{L} \d x \int_0^{L}\d x^\prime \int_{-\frac{\mathrm{min}\left(W(x),W(x')\right)}{2}}^{\frac{\mathrm{min}\left(W(x),W(x')\right)}{2}} \d y \\
& \times E(x-x_0,y,w_0)  E(x^\prime-x_0,y,w_0) \\
& \times \frac{\phi_n(x)-\phi_n(x^\prime)}{x-x^\prime}. 
\end{split}
\end{equation}
Similarly, the sum of all the photodiodes $S$ can be deduced from Eq.~\ref{eq.Dx1}:
\begin{equation}
\label{eq.SS}
\begin{split}
& S = \int_0^{L} \d x \int_0^{L}\d x^\prime \int_{-\frac{W(x)}{2}}^{\frac{W(x)}{2}} \d y \int_{-\frac{W(x^\prime)}{2}}^{\frac{W(x^\prime)}{2}} \d y^\prime\\
& \times  E(x-x_0,y,w_0)  E(x^\prime-x_0,y^\prime,w_0) \\
& \times e^{2ik(\delta(x,y)-\delta(x^\prime,y^\prime))} \frac{k}{2\pi F} \int_{-\infty}^{+\infty} \dX e^{-ikX(x-x^\prime)/F} \\
& \times \frac{k}{2\pi F} \int_{-\infty}^{+\infty} \dY e^{-ikY(y-y^\prime)/F}.
\end{split}
\end{equation}
In this case the $0^\mathrm{th}$ order of the displacement contribution is non-zero due to the integral being symmetric, thus $S$ is independent of the mode considered, as expected. Expressing the Delta functions, Eq.~\ref{eq.SS} simplifies:
\begin{equation}
\label{eq.S}
\begin{split}
S = \int_0^{L} \d x \int_{-\frac{W(x)}{2}}^{\frac{W(x)}{2}} \dy E(x-x_0,y,w_0)^2
\end{split}
\end{equation}
From Eq.s \ref{eq.Dn} and \ref{eq.S} it is then possible to express the contrast $C_n = D_n/S$:
\begin{equation}
\label{eq.Cxx}
\begin{split}
C_n = & \,  \frac{4\delta_n}{\lambda S} \int_0^{L} \d x \int_0^{L}\d x^\prime \int_{-\frac{W(x)}{2}}^{\frac{W(x)}{2}} \d y \\
& \times E(x-x_0,y,w_0)  E(x^\prime-x_0,y,w_0) \\
& \times \frac{\phi_n(x)-\phi_n(x^\prime)}{x-x^\prime}. \\
 = & \, \sigma_n(x_0,w_0)\delta_n, 
\end{split}
\end{equation}
which gives the sensitivity of Eq.~\ref{eq.Cequalsigmadelta}. 

In the same way, we can recover the torsional sensitivity. From the second of Eqs.~\ref{eq.Cx2}, the contrast $C_m$ is the difference between the upper and lower quadrants. Following the procedure used to retrieve $C_n$, we consider the torsion mode $m$ in Eq.~\ref{eq.deltam}: $\delta(x,y) = \theta_m y \phi_m(x)$. The difference between the left and right quadrants is:
\begin{equation}
\begin{split}
D_m & =  \frac{4 \theta_m}{\lambda } \int_0^{L} \dx \phi_m(x) \\
& \times \left|\int_{-\frac{W(x)}{2}}^{\frac{W(x)}{2}} \dy E(x-x_0,y,w_0)\right|^2.
\end{split}
\end{equation}\newpage
\noindent The contrast then reads:
\begin{equation}
\begin{split}
C_m = \,& \frac{4 \theta_m}{\lambda S} \int_0^{L} \dx \phi_m(x) \\
& \times \left|\int_{-\frac{W(x)}{2}}^{\frac{W(x)}{2}} \dy E(x-x_0,y,w_0)\right|^2 \\
 = &\, \sigma_m(x_0,w_0) \theta_m, 
\end{split}
\end{equation}
which gives the torsional sensitivity of Eq.~\ref{eq.sm}.
\vfill

\bibliography{MultimodeOBD}

\end{document}